\documentclass[12pt]{amsart}
\usepackage{graphicx}
\usepackage{amscd}
\usepackage{latexsym}


\newcommand{\emb}{\hookrightarrow}

\newcommand{\field}[1]{\mathbb{#1}}
\newcommand{\rz}{\field{R}}
\newcommand{\cz}{\field{C}}
\newcommand{\zz}{\field{Z}}
\newcommand{\nz}{\field{N}}

\newcommand{\ti}{\tilde  }

\newcommand{\place}{\hspace{2mm} \cdot \hspace{2mm}}
\newcommand{\spc}{$\textrm{Spin}^\cz$}
\newcommand{\cp}{\mathbb{CP}^2}

\newtheorem{theorem}{Theorem}[section]
\newtheorem{definition}[theorem]{Definition}

\newtheorem{lem}[theorem]{Lemma}

\newtheorem{rem}[theorem]{Remark}

\title{Towards a Nonperturbative Covariant Regularization in 4D Quantum Field
  Theory}

\begin{document}

\footnotetext[1]{part of project P11783-PHY of the ''Fonds zur F\"{o}rderung der 
 wissenschaftlichen Forschung in \"{O}sterreich''} 

\maketitle

\begin{center}
H. Grosse$^{1*}$, A. Strohmaier$^{**}$\\
\tiny
$^{*}$Institute for Theoretical Physics, University of Vienna, Boltzmanngasse 5,
A-1090 Vienna, Austria \\
e-mail: grosse@doppler.thp.univie.ac.at\\
$^{**}$Institute for Theoretical Physics, TU- Graz, Petersgasse 16,
A-8010 Graz, Austria \\
e-mail: alexan@itp.tu-graz.ac.at
\end{center}

\begin{abstract}
  We give a noncommutative version of the complex projective
  space $\cp$ and show that scalar QFT on this space is free
  of UV divergencies. The tools necessary to investigate
  Quantum fields on this fuzzy $\cp$ are developed and several
  possibilities to introduce spinors and Dirac operators
  are discussed.
\end{abstract}

\normalsize 


\section{Introduction}

Recently methods of noncommutative geometry (\cite{Connes}) were used
to introduce a covariant regularization procedure
for Quantum fields on the 2 sphere (\cite{Gross1,Gross2}).
There is hope that with these methods there will be
a nonperturbative understanding of quantum effects. For a treatment
of the chiral anomaly in the Schwinger model see ~\cite{GrossSchw}.

The idea is to approximate the algebra of functions on
the space by matrix algebras and to encode the geometrical
information of the space in these algebras.
In this way it becomes possible to construct models
of QFT on the virtual spaces, the matrix algebras are thought
to be functions on. 
Since the theory has only finite degrees of freedom the
problems one usually deals with in QFT are absent.
For examples of these matrix geometries see 
~\cite{MadFuzzy,MadIntr,Gross}.

We will show that the $\cp$, as a 4 dimensional Riemannian
manifold on which the group $SU(3)$ acts isometrically,
can by treated in this way.
We construct the Laplace operator on the fuzzy $\cp$
and show that the spectrum is the same as in the
classical case, except for a truncation at higher
eigenvalues, which can be interpreted as a UV cutoff.
Since $SU(3)$ acts on the matrix algebras by automorphisms
we did not lose the $SU(3)$ covariance by this procedure.

We discuss the quantization of several vector bundles,
among them a family of \spc-bundles, and compare the structures
with the classical case.
As well as for the functions we find that in the quantized
sections higher order representations of $G$ are missing.
A straightforward generalization of the classical \spc-Dirac
operators is introduced and the spectra are calculated.
It turns out, that zero modes and index are different from
the commutative case.
A possible solution to this problem is presented.

\section{The classical case}

We start with a group theoretic construction of
the classical structures on the $\cp$.
Let $G^\cz$ be the Lie group $SL_3(\cz)$ and
$g^\cz$ its Lie algebra. We choose a Cartan subalgebra
and a corresponding system of roots $\Delta$, positive
roots $\Delta^+$, fundamental roots $\Delta^f$, and use the
standard Cartan-Weyl basis
$ \lbrace H_\delta \rbrace_{\delta \in \Delta^f} \cup
\lbrace X_\delta, X_{-\delta} \rbrace_{\delta \in \Delta^+}$.
$H_\delta$ is the root vector and $X_\delta$ the root element
corresponding to the root $\delta$.
\begin{figure}[h]\centering
\includegraphics*[width=70mm]{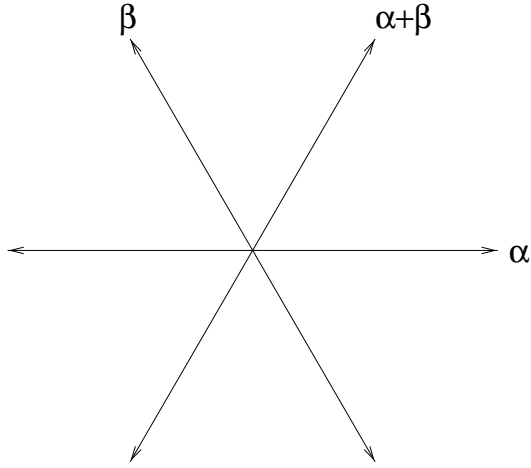}
\caption{Root Diagram of the \(A_{2}\)} \label{roots}
\end{figure}
$\Delta^f$ consists of two elements $\alpha$ and $\beta$ (see Figure ~\ref{roots}).
Hence we have
\begin{equation}
g^\cz = \mathrm{span}_\cz \lbrace H_{\alpha},H_{\beta},X_{\alpha},
X_{\beta},X_{-\alpha},X_{-\beta},X_{\alpha+\beta},X_{-(\alpha+\beta)} \rbrace
\end{equation}
The compact real form $g \subset g^\cz$ is
\begin{eqnarray}
g=\mathrm{span}_\rz \lbrace i H_{\alpha},i H_{\beta},X_{\alpha}+
X_{-\alpha},X_{\beta}+X_{-\beta},X_{\alpha+\beta}+X_{-(\alpha+\beta)},\\
i(X_{\alpha}-X_{-\alpha}),i(X_{\beta}-X_{-\beta}),
i(X_{\alpha+\beta}-X_{-(\alpha+\beta)})
\rbrace \nonumber
\end{eqnarray}
and the corresponding connected subgroup $G \subset G^\cz$ is the
group $SU(3)$.
The two fundamental weights are
\begin{equation}
\lambda_{1} = \frac{1}{3}(2 \alpha + \beta) \qquad
\lambda_{2} = \frac{1}{3}(\alpha + 2 \beta)
\end{equation}
We denote the Verma module with highest weight
$\lambda=n_{1}\lambda_{1}+n_{2}\lambda_{2}$ by $D(n_1,n_2)$
the fundamental roots chosen such that $D(1,0)$ corresponds to the
fundamental representation of the $SU(3)$.
The dimensions of these representations are given by
\begin{equation}
dim(D(n_{1},n_{2}))=\frac{(n_{1}+1)(n_{2}+1)(n_{1}+n_{2}+2)}{2}
\end{equation}
and the image of the quadratic Casimir element
\begin{equation}
\mathbf{C}=\sum_{\delta \in \Delta^+} (X_{\delta}X_{-\delta}+
  X_{\delta}X_{-\delta})+\sum_{\delta \in \Delta^f} \frac{1}{2} H_{\delta}^2
\end{equation}
is the multiplication operator by the number
\begin{equation}
\mathbf{C}(n_1,n_2)=
\frac{2}{3}(n_1^2+n_2^2+n_1 n_2+3(n_1+n_2))
\end{equation}
To each Verma module there corresponds an (up to equivalence unique)
unitary irreducible representation of $G$ and a holomorphic irreducible
representation of $G^\cz$, which we denote by the same symbol.

Let $p \subset g^\cz$ now be the parabolic subalgebra
\begin{equation}
p= \mathrm{span}_\cz\lbrace H_\alpha, H_\beta, X_\alpha,
X_{-\alpha}, X_{-\beta}, X_{-(\alpha+\beta)}
\rbrace
\end{equation}
and $P$ the analytic subgroup of $G^\cz$ with Lie algebra
$p$.
The quotient space $M:=G^\cz/P$ is a complex flag manifold
(see e.g. ~\cite{Wolf,Wallach,Warner})
and has a compact realization $M=G/K$ where $K$ is the
subgroup of $G$ with Lie algebra $k:=g \cap p$.
In our case this is the group $S(U(2) \times U(1)) \subset SU(3)$.
This is just the subgroup that leaves the lowest weight subspace
of $D(1,0)$ invariant, and our
space $M$ is just the complex projective space $\cp$, the
space of comlex lines in $\cz^3$.

$M$ is in a canonical way a K\"{a}hler manifold.
The action of $G$ on $M$ is holomorphic and leaves the K\"{a}hler
structure invariant.
There exists a unique normalized $G$-left invariant measure on
$M$ which we denote by $\mu$.

\subsection{The quasi-left regular representation}

In this section we give the decomposition of the unitary
representation of $G$ on the space $\mathrm{L}^2(M,\mu)$,
given by the action
\begin{equation}
 g f(x) = f(g^{-1} x)
\end{equation}
For the decomposition we use the fact that $M$ is a Riemannian
symmetric space of compact type (\cite{Helg,Helg2}). This means that there is
a real form $G^\rz \subset G^\cz$ with Iwasawa decomposition
$G^\rz=KAN$, where $A$ is the abelian and $N$ the nilpotent part.
We denote by $g^\rz,a,n \subset g^\cz$ the corresponding real
Lie algebras.
$a$ can be extended to a maximal abelian Lie subalgebra $h$ of $g^\rz$
such that its complexification $h^\cz$ is a Cartan subalgebra of $g^\cz$.
The weights are elements of the dual of the vectorspace $h_\rz=i(h \cap
k)+a$.
We choose again fundamental roots and denote by $\Sigma^+$ the
positive roots that do not vanish on $a$ and interpret them as
linear forms on $a$. The Killing form induces a scalar product
$\langle \cdot \vert \cdot \rangle$
on the dual $a^*$.
We have
\begin{theorem}[Helgason]
  A representation with highest weight $\lambda$
  contains a $K$-invariant vector if and only if
  \begin{enumerate}
     \item \(\lambda(i(h \cap k))=0\)
     \item \(\frac{\langle\lambda \vert \delta\rangle}
      {\langle \delta \vert \delta \rangle} \in 
      \nz_{0} \quad \forall \delta \in \Sigma^{+}\)
\end{enumerate}
\end{theorem}
(see ~\cite{Helg2})

In our case a short calculation shows that one can choose
another Cartan subalgebra such that
$$
k=g \cap \mathrm{span}_{\cz} \lbrace H_{\beta}-H_{\alpha},
X_{\beta}+X_{-\alpha},X_{-\beta}+X_{\alpha},
X_{\alpha+\beta}+X_{-(\alpha+\beta)} \rbrace
$$
and with
\begin{eqnarray}
 a=\mathrm{span}_{\rz}\lbrace H_{\alpha}+H_{\beta}\rbrace \hspace{2cm}\nonumber\\
 n=\mathrm{span}_{\rz}\lbrace (X_{\alpha}-X_{-\alpha}),
(X_{\beta}-X_{-\beta}),(X_{\alpha+\beta}-X_{-(\alpha+\beta)}) \rbrace\nonumber
\end{eqnarray}
$g^\rz=k+a+n$ is a real form of $g^\cz$.
One obtains that the representations of $G$ that contain a $K$ invariant
vector are just the representations with highest weight
\begin{equation}
\lambda=n(\lambda_{1}+\lambda_{2}) \quad n \in \nz_{0}
\end{equation}

The space of $K$-invariant vectors in such a representation is
furthermore one dimensional and by Frobenius theorem exactly
those representations occur with multiplicity one in the
quasi-left regular representation \footnote{the quasi-left regular
  representation
  on $G/K$ is just the representation of $G$ induced by the trivial
  representation of $K$}.

Hence we get for our case the following result
\begin{theorem}\label{decomp}
  Let $M=\cp$ with the canonical $SU(3)$ action.
  As $SU(3)$-modules we have
  $$\mathrm{L}^2(M,\mu) \cong \bigoplus_{n=0}^{\infty} D(n,n)$$
and the subspace of $\mathrm{L}^2(M,\mu)$ which is, as an SU(3)-module,
isomorphic to $D(n,n)$ is just the space of functions of the form
$$ M \to \cz, \quad uK \to \langle u e_n \vert v \rangle, \quad v \in
D(n,n)$$
where $e_n \in D(n,n)$ is an (up to a scalar unique) K-invariant vector.
\end{theorem}

\subsection{The Laplace operator}
Identifying the Lie algebra $g^\cz$ with the complex left invariant vector
fields on the group $G$ and those with differential operators
on $C^\infty(G)$, we get a natural action of the universal
enveloping algebra $\mathcal{U}(g^\cz)$ on $C^\infty(G)$.
The quadratic Casimir element $\mathbf{C}$ becomes is a biinvariant
Differential operator $\Delta_G$ of second order and is equal to the
Laplace-Beltrami operator of the (up to a scalar unique) biinvariant 
metric on $G$.
Thus it can be restricted to the $K$-right invariant functions on $G$
and we get a differential operator $\Delta_M: C^\infty(M) \to  C^\infty(M)$.
This operator coincides with the metric Laplacian on $M$.
We get
\begin{theorem}
 The Laplace operator $\Delta_M$ on the $\cp$
 is essentially self-adjoint on  $C^\infty(\cp)$. The
 spectrum of its closure is purely discrete and the
 eigenvalues $\lambda$ and multiplicities $\mathrm{mult}_\lambda$ are
 \begin{eqnarray}
   \lambda_n &=& 2n(n+1), \quad n \in \nz_0 \nonumber \\
   \mathrm{mult}_{\lambda_n} &=& (n+1)^3 \qquad \nonumber
 \end{eqnarray}
\end{theorem}
Proof:
We restrict the domain to the finite linear
span of vectors in the irreducible subspaces of the group action,
on each of which $\Delta_M$ acts by multiplication
with $2n(n+1)$. Hence the domain contains a total set of analytic
vectors and we found the operator in its spectral decomposition.
\hfill \(\Box\)

\subsection{Homogeneous vector bundles over $M$}
The homogeneous vector bundles over $M$ are in one to one
correspondence with the finite dimensional representations
of the group $K$.
If $V$ is a homogeneous vector bundle over $M$ and $\tau$ the
representation of $K$ at the fibre $V_{eK}$ at the identity class $eK$,
then $V$ is isomorphic as a G-vector bundle to the associated
bundle $G \times_\tau V_{eK}$. The sections of this bundle
can be identified with the equivariant $V_{eK}$-valued functions
on $G$, that is the functions $f$ that satisfy the equivariance
condition $f(xk)=\tau(k^{-1})f(x) \quad \forall x \in G, k \in K$.

The homogeneous vector bundles over $M$ have a canonical holomorphic
structure since $\tau$ can be extended to a holomorphic representation
$\tilde\tau$ of $P$ and the vector bundle can be constructed as
$G^\cz \times_{\tilde\tau} V_{eK}$ .

We define several useful bundles on $M$.
Let $\tau(n)$ for $n \in \zz$ be the irreducible (one dimensional)
representation of $K$ with highest weight $-n\lambda_2$.
We denote the homogeneous line bundle $G \times_{\tau(n)} \cz$
by $L^n$. For example $L$ is the hyperplane bundle and $L^{-1}$
the tautological line bundle over $M$.

Let $\chi$ be the representation of $K$ with highest weight $\lambda_1$
and $\chi^*$ the dual representation, we define $H$ and $H^{-1}$ to be the
bundles $G \times_{\chi} \cz^2$ and $G \times_{\chi^*} \cz^2$ respectively.

Since the irreducible representations of $K$ are unitary these
vector bundles have natural hermitian structures, and we
denote by $\Gamma^2(V)$ the Hilbert space of square integrable 
sections and by $\Gamma_{hol}(V)$ the finite dimensional Hilbert space
of holomorphic sections of the bundle $V$.
The representation of $G$ on $\Gamma^2(G \times_\tau V_\tau)$ is
just the representation induced by $\tau$.
\begin{theorem}
For the bundles $L^n$ we have the following isomorphisms of $G$-modules
\begin{enumerate}
\item for \(n \geq 0\): \(\Gamma^2(L^n) \cong
  \bigoplus_{m=0}^{\infty} D(m,m+n)\)
\item for \(n < 0\): \(\Gamma^2(L^n) \cong
  \bigoplus_{m=0}^{\infty} D(m-n,m)\)
\end{enumerate}
\end{theorem}

Sketch of Proof:
We specialize to $n \geq 0$ since the proof is analogous for the
other case.
The sections of $L^n$ will be identified with the equivariant functions
on $G$. For each section $f$ there is a family of $K$-right invariant
functions $(f_i)_{i=1 \ldots \mathrm{dim}(D(0,n))}$ on $G$, such that
$$
 f(x)=\sum_{i=1}^{\mathrm{dim}(D(0,n))} f_i(x) \langle x h_n \vert
 v_i \rangle
$$
where $h_n$ is a nonzero highest weight vector and
$\lbrace v_i \rbrace_{i=1 \ldots \mathrm{dim}(D(0,n))}$ an orthonormal
basis in $D(0,n)$. To see this simply choose
\(f_i(x)=f(x)\langle v_i \vert x h_n \rangle\).
Hence using theorem ~\ref{decomp} the subspace
\begin{displaymath}
\mathcal{W}:=\mathrm{span}_\cz\lbrace \langle \place e_m \vert u^{(m)}
\rangle \langle \place h_n \vert v \rangle: u^{(m)} \in D(m,m),
v \in D(0,n), m \in \nz_0 \rbrace
\end{displaymath}
is dense in $\Gamma^2(L^n)$.
The subspaces
\begin{displaymath}
\mathcal{W}^N:=\mathrm{span}_\cz\lbrace \langle \place e_m \vert u^{(m)}
\rangle \langle \place h_n \vert v \rangle: u^{(m)} \in D(m,m),
v \in D(0,n), m \leq N \rbrace
\end{displaymath}
define a filtration of $\mathcal{W}$.
A short calculation using $D(0,m) \otimes D(m,0)=\oplus_{i=0}^m D(i,i)$ gives
\begin{displaymath}
\mathcal{W}^N=\mathrm{span}_\cz\lbrace \langle \place h_{n+m} \vert u^{(m+n)}
\rangle \langle \place l_m \vert v^{(m)} \rangle: u^{(m)} \in D(0,n+m),
v^{(m)} \in D(m,0), m \leq N \rbrace
\end{displaymath}
where $l_m$ is a nonzero lowest weight vector in $D(m,0)$.
Since $D(0,m+n) \otimes D(m,0)=\oplus_{i=0}^m D(i,i+n)$
and the vector $h_{n+m} \otimes l_m \in D(0,n+m) \otimes D(m,0)$ is cyclic,
we can choose a family of nonzero vectors $e^m \in D(m,m+n)$ such that
$$
\mathcal{W}^N=
\mathrm{span}_\cz \lbrace
\langle \place e^m \vert v^{(m)} \rangle : v^{(m)} \in D(m,m+n),
m \leq N \rbrace
$$
and therefore
$$
\mathcal{W}=
\mathrm{span}_\cz \lbrace
\langle \place e^m \vert v^{(m)} \rangle : v^{(m)} \in D(m,m+n),
m \in \nz_0 \rbrace
$$
This establishes the desired isomorphism of modules. \hfill \(\Box\)

The same method allows us to decompose the representation
on the space of sections of another class of bundles
\begin{theorem}
For the bundles $L^n \otimes H^{-1}$ we have the following isomorphisms of $G$-modules
\begin{enumerate}
\item for \(n \geq 0\): \(\Gamma^2(L^n \otimes H^{-1}) \cong
     \bigoplus_{m=0}^\infty (D(m,m+n+2) \oplus D(m+1,m+n))\)
   \item for \(n<0\): \(\Gamma^2(L^n \otimes H^{-1}) \cong
     \bigoplus_{m=0}^\infty (D(m-n-1,m+1) \oplus D(m-n+1,m))\)
\end{enumerate}
\end{theorem}

Since $G$ acts on $M$ by holomorphic automorphisms
the bundles of antiholomorphic $k$-forms $\Lambda^{0,k}$ are homogeneous
vector bundles. The vector space of antiholomorphic tangent vectors at
$eK$ can be identified with the nilpotent radical $u$ of the parabolic
subalgebra $p \subset g^\cz$, which is just
$\mathrm{span}_\cz \lbrace X_{-(\alpha+\beta)},X_{-\beta} \rbrace$.
The isotropy representation is the adjoint action of $K$ on $u$.
Taking the dual representations this leads to the following
isomorphisms of $G$-bundles.
\begin{eqnarray}
 \Lambda^{0,0} &\cong& L^0 \nonumber\\
 \Lambda^{0,1} &\cong& L \otimes H^{-1}\\
 \Lambda^{0,2} &\cong& L^3 \nonumber
\end{eqnarray}

\subsection{\spc-bundles}
It is well known that the $\cp$ does not admit a Spin-structure.
It does however admit a \spc-structure and we give here a family
of \spc-bundles.
Since $M$ is a K\"{a}hler manifold the bundle of antiholomorphic
forms $\Lambda^{0,*}$ with the natural $\zz_2$-grading is a canonical
\spc-bundle and the Dolbeault-Dirac operator
$D=\sqrt{2} (\overline\partial+\overline\partial^*)$ is a \spc-Dirac operator
(see e.g. ~\cite{Lawson}).
Tensoring this bundle with some line bundle we obtain other \spc-bundles.

Tensoring the canonical \spc-bundle $S$ with the line bundle $L^m$
and using the above described isomorphisms of $G$-bundles
we get the \spc-bundles $S_m$:
\begin{eqnarray}
S^+_m & \cong & L^m \oplus L^{m+3} \quad \ldots \textrm{even part}\nonumber \\
S^-_m & \cong & L^{m+1} \otimes H^{-1} \quad \ldots \textrm{odd part}\\
S_m & = & S^+_m \oplus S^-_m \nonumber
\end{eqnarray}
The sections of $S_m$ are identified with the $\cz^4$-valued functions $\phi$
on $G$ which satisfy the equivariance condition
\begin{displaymath}
\left( \begin{array}{c} \phi_1(gk) \\ \phi_2(gk) \\ \phi_3(gk)
    \\ \phi_4(gk) \\ \end{array} \right)= \tau(-m)
\left( \begin{array}{cccc}
    
   1 & 0 & 0 & 0\\
   0 & \tau(-3)(k) & 0 & 0 \\
   0 & 0 & \tau(-1)(k) \chi_{11}(k) & \tau(-1)(k) \chi_{12}(k)\\
   0 & 0 & \tau(-1)(k) \chi_{21}(k) & \tau(-1)(k) \chi_{22}(k)\\
    
\end{array}
\right)
\left( \begin{array}{c} \phi_1(g) \\ \phi_2(g) \\ \phi_3(g) \\ \phi_4(g) \\
\end{array} \right)
\end{displaymath}
for all \(k \in K, g \in G\). Here the $\chi_{ik}$ are the matrix elements
of the representation $\chi$.
The basis in the representation space of $\chi$ can be chosen in such
a way that the Dolbeault-Dirac operator $D_m$ takes the form
\begin{equation}
 D_m=\sqrt{2} 
 \left( \begin{array}{cccc}
   0 & 0 & \ti X_{\alpha+\beta} & \ti X_\beta \\
   0 & 0 & -\ti X_{-\beta} & \ti X_{-(\alpha+\beta)} \\
   \ti X_{-(\alpha+\beta)} & -\ti X_\beta & 0 & 0\\
   \ti X_{-\beta} &  \ti X_{\alpha+\beta} & 0 & 0\\   
\end{array}
\right)=
\left( \begin{array}{cc}
    0 & \ti D^*\\
    \ti D & 0 \\
\end{array} \right)
\end{equation}
where $\ti X_\delta$ is the differential operator corresponding
to the left invariant vector field $X_\delta$ on $G$.
This is an example of an abstract supersymmetric Dirac operator.
We have
\begin{theorem}
The \spc-Dirac operator \(D_m\) is essentially self-adjoint on the
space of smooth sections of $S_m$. The spectrum of its closure is purely
discrete. The eigenvalues \(\lambda\) and multiplicities \(\mathrm{mult}_\lambda\)
are:
\begin{enumerate}
  \item \(\lambda_n^\pm=\pm \sqrt{2n(n+2)+2\vert m\vert(n+1)-2m} \quad n \in
    \nz^+_0\)\\
    \(\mathrm{mult}_{\lambda_n^\pm}=(n+\frac{\vert m \vert +2}{2})(n+\vert m
    \vert +1)(n+1)\)
  \item \(\lambda_n^\pm=\pm \sqrt{2n(n+2)+2\vert m+3\vert(n+1)+2(m+3)} \quad n \in
    \nz^+_0\)\\
    \(\mathrm{mult}_{\lambda_n^\pm}=(n+\frac{\vert m+3 \vert +2}{2})(n+\vert m+3
    \vert +1)(n+1)\)
\end{enumerate}
The index is \(\mathrm{index}(D_m)=\mathrm{mult}_{\lambda=0}=
\frac{(m+1)(m+2)}{2}\)
\end{theorem}
Sketch of Proof: The essential self-adjointness follows from Nelsons
trick. Using the transformation properties on the group an easy
calculation shows
\begin{displaymath}
 \ti D_m^* \ti D_m= 
 \left( \begin{array}{cc}
   \Delta_G-2m-\frac{2}{3}m^2 & 0 \\
   0 & \Delta_G-2m-\frac{2}{3}m^2 \\
\end{array}
\right)
\end{displaymath}
The abstract Foldy-Wouthuysen transformation and the decompositions
of the $G$-modules $\Gamma(L^m)$, $\Gamma(L^{m+3})$, and
$\Gamma(L^{m+1} \otimes H^{-1})$ gives the desired result.\hfill \(\Box\)

\section{Quantization of the $\cp$}

Since $M$ is a compact K\"{a}hler manifold and $L$ a Quantum
line bundle we can use the Berezin-Toeplitz quantization
procedure (see e.g. ~\cite{Schlich}).
We fix an $N \in \nz_0$ and let $\Pi_N$ be the orthogonal projection 
onto the subspace
$\mathcal{H}^N:=\Gamma_{hol}(L^N) \subset \Gamma^2(L^N)$.
For each \(f \in C^\infty(M)\) we get an operator \(T^N(f) \in
\mathrm{End}(\mathcal{H}^N)\), defined by
\begin{displaymath}
 T^N(f):=\Pi_N M_f \Pi_N
\end{displaymath}
where \(M_f\) is the multiplication-operator associated with
\(f\).
The corresponding surjective map \(C^\infty(M) \to
\mathrm{End}(\mathcal{H}^N)\) is called Toeplitz quantization map.
The algebra \(\mathcal{A}_N:=\mathrm{End}(\mathcal{H}^N)\)
is the algebra of quantized functions, which for
high values of \(N\) approximates the algebra \(C^\infty(M)\)
in a certain sense.

By the Bott-Borel-Weyl theorem (see ~\cite{Wallach, Warner})
we have the isomorphism
of $G$-modules $\mathcal{H}^N \cong D(0,N)$.
Identifying these two spaces we obtain for the
quantized algebra of functions the matrix algebra
\begin{equation}
  \mathcal{A}_N=Mat(\frac{(N+1)(N+2)}{2},\cz)
\end{equation}
on which $G$ acts by inner automorphisms
\begin{equation}
 G \times \mathcal{A}_N \to \mathcal{A}_N: \quad
   (g,a) \to \pi(g) a \pi^{-1}(g)
\end{equation}
where $\pi$ is the representation which corresponds to
the $G$-module $D(0,N)$.
If $ \tilde P_N$ is the orthogonal projection onto the highest weight subspace
of $\mathcal{H}^N$ we define the coherent state function on $M$
by
\begin{equation}
 P_N : M \to \mathrm{End}(\mathcal{H}^N), \quad gK \to \pi(g) \ti P_N \pi^{-1}(g)
\end{equation}
and we get for the Toeplitz quantization map
\begin{equation}
  T^N_f=\frac{(N+1)(N+2)}{2} \int_{M} f(x) P_N(x) d\mu(x)
\end{equation}
which is equivariant, real, positive, and norm decreasing.

Moreover the following results hold
\begin{theorem} \hfill
\begin{enumerate}
\item The restriction \(t^N:=T^N\vert_{\mathcal{E}^N}\) ist bijective.
\item \(\lim_{N\to\infty}(t^N)^{-1}T^N(f)=f\) in the uniform topology
on \(C^\infty(M)\)
\item for \(f_1 \ldots f_n \in C^\infty(M)\) we have
with the \(C^\ast\)-norms on the \(\mathcal{A}_N\):
\begin{displaymath}
\Vert T^N_{f_1} \cdots  T^N_{f_n} - T^N_{f_1 \cdots f_n} \Vert
=\mathcal{O}(\frac{1}{N}) \quad \mathrm{for} \quad N \rightarrow \infty
\end{displaymath}
\item If \(\mathrm{tr}_1(\cdot)\) is the normalized trace on
\(\mathcal{A}_N\), we have:
\begin{displaymath}
\int_{M} f(x) d\mu(x) = \mathrm{tr}_1(T^N_f) \quad 
\forall f \in C^\infty(M), N \in \nz_0
\end{displaymath}
\item
for \(f_1 \ldots f_n \in C^\infty(M)\) holds:
\begin{displaymath}
\int_{M} f_1(x) \cdots f_n(x) d\mu(x) = 
\mathrm{tr}_1(T^N_{f_1} \cdots T^N_{f_n})+\mathcal{O}(\frac{1}{N}) \quad 
\mathrm{for} \quad N \rightarrow \infty
\end{displaymath}
\item
  $\Vert T^N_f \Vert = \Vert f \Vert_{\mathrm{sup}} +
  \mathcal{O}(\frac{1}{N}) \quad \mathrm{for} \quad N \rightarrow \infty$
\end{enumerate}
\end{theorem}
Here $\mathcal{E}^N \subset C^\infty(M)$ is the space of
truncated functions, that is the subspace that contains
only the $G$-representations $D(n,n)$ for $n \leq N$. 

It follows that as $G$-modules $\mathcal{A}_N$ and
$\oplus_{n=0}^{N} D(n,n)$ are isomorphic. This is a
kind of cut-off automated by the quantization.
One introduces now the linear inclusions
\begin{equation}
i_N: \mathcal{A}_N \emb \mathcal{A}_{N+1}, \quad a \rightarrow 
t^{N+1}(t^N)^{-1}a
\end{equation}
and the sequence
\begin{equation}
(A_{\ast},i_{\ast}):=
A_0 \stackrel{i_1}{\emb} \ldots \stackrel{i_{N-1}}{\emb}  \mathcal{A}_N 
\stackrel{i_N}{\emb}  \mathcal{A}_{N+1}  \stackrel{i_{N+1}}{\emb} \ldots
\end{equation}
The vector space limit of this sequence is in a canonical
way a normed space and its closure has the structure of
a $C^*$-algebra, the product being the limit $N \to \infty$
of the products in the single algebras.
This algebra can be shown to be isomorphic
to the algebra of continuous functions on $M$. Moreover
the isomorphism can be chosen to be equivariant.
In this sense the sequence approximates the $\cp$.
We therefore understand the algebras $\mathcal{A}_N$
as the algebra of functions on the ''virtual'' fuzzy $\cp$ of
order $N$.
This requires however to have as a ''correspondence principle''
the Toeplitz quantization maps or equivalently the above
sequence in mind.

\section{The Laplace operator on the fuzzy $\cp$}

Since $G$ acts on $\mathcal{A}_N$ by inner automorphisms we get by
differentiation an action of the Lie algebra $g^\cz=\mathrm{sl}_3(\cz)$
by inner derivations. If $\pi$ is the representation of $g^\cz$
which corresponds to the $g^\cz$-module $D(0,N)$ this action
reads as follows
\begin{equation}
  x a = [\pi(x),a] \quad x \in g^\cz, a \in \mathcal{A}_N
\end{equation}
This gives a representation of the universal enveloping algebra
$\mathcal{U}(g^\cz)$ and we define in analogy to the commutative
case the Laplace operator
\begin{equation}
\Delta_N: \mathcal{A}_N \to \mathcal{A}_N
\end{equation}
to be the image of the quadratic Casimir element,
or more explicitly:
\begin{equation}
 \Delta_N a:=\left(\sum_{\delta \in \Delta^+} (X_{\delta}X_{-\delta}+
  X_{\delta}X_{-\delta})+\sum_{\delta \in \Delta^f} \frac{1}{2} H_{\delta}^2
  \right) a
\end{equation}
Since the quantization map is equivariant we have
\begin{equation}
 \Delta_N(a) = t^N \left(\Delta_M (t^N)^{-1}(a)\right)
\end{equation}
Using the decomposition of the $g^\cz$-module $\mathcal{A}_N$
one gets easily
\begin{theorem}
The eigenvalues \(\lambda\) of the Laplacian \(\Delta_N\) and their
 multiplicities \(\mathrm{mult}_\lambda\) are:
 \begin{displaymath}
  \lambda_n=2n(n+2), \quad \mathrm{mult}_{\lambda_n}=(n+1)^3 \quad n \in 
  \lbrace 0, \ldots , N \rbrace 
 \end{displaymath}
\end{theorem}

The spectrum shows the expected UV cutoff.

The above action of $g^\cz$ on $\mathcal{A}_N$ can be thought of
as a Lie algbera homomorphism from $g^\cz$ to the Lie algebra
of derivations $\mathrm{Der}(\mathcal{A}_N)$ of the algebra. 
The image, as a subspace of $\mathrm{Der}(\mathcal{A}_N)$, gives rise
to a derivation based differential culculus on $A_N$ (see ~\cite{Dubvio}).
With the natural scalar product on the one-forms the
Laplace operator has the natural form $\Delta_N=d^*d$, as in the
commutative case.

\section{Quantization of homogeneous vector bundles over $\cp$}

Vector bundles in the noncommutative case should be
finitly generated projective modules over the algebra.
In ~\cite{Eli2} the concept of Toeplitz quantization
was generalized to the quantization of bundles.

Let \(E\) be a holomorphic hermitian vector bundle with base \(M\) and
\(E^*\) be the dual bundle.
Let furthermore \(\mathcal{H}_E^N\) be the Hilbert space
of holomorphic sections of the bundle \(E^* \otimes L^N\), that is
\(\mathcal{H}_E^N:=\Gamma_{hol}(E^* \otimes L^N)\).
Using the orthogonal projection
\begin{equation}
 \Pi_N^E \in \mathcal{B}(\mathrm{L}^2(M,E^* \otimes L^N))
\end{equation}
onto \(\mathcal{H}_E^N\), we can define the surjective map
\begin{equation}
  T_E^N : \quad \Gamma(M,E) \to \mathrm{Hom}_\cz(\mathcal{H}_E^N,\mathcal{H}^N), \quad
  \nu \to \Pi_N \nu \Pi_N^E
\end{equation}
According to ~\cite{Eli} this gives the quantization of the bundle
\(E\) in the following sense
\begin{equation}
 \lim_{N \to \infty} \Vert T^N(f) T^N_E(\nu) - T^N_E(f\nu) \Vert = 0 
\end{equation}

Clearly \(\mathcal{M}_N^E:=\mathrm{Hom}_\cz(\mathcal{H}_E^N,\mathcal{H}^N)\)
is a finitely generated projective left \(\mathcal{A}_N\)-module.

Let $\pi$ be an irreducible representation of $K$ and \(E\)
the associated homogeneous vector bundle over \(M\).
If \(\pi\) has highest weight \(n_1\lambda_1+n_2\lambda_2\),
the isotropy representation of \(E^*\otimes L^N\) has
highest weight
\begin{equation}
  \gamma=n_1\lambda_1+(N-n_1-n_2)\lambda_2
\end{equation}
Then by the Bott-Borel-Weyl theorem
\begin{equation}
  \Gamma_{hol}(E^* \otimes L^N) \cong D(n_1,N-n_1-n_2)
\end{equation} as G-modules,
if we set by definition \(D(n_1,n_2)=\lbrace 0 \rbrace\)
if \(n_1 < 0\) or \(n_2 < 0\).

\subsection{The Bundles $E=L^n$}

The isotropy representation of \(L^n\) was \(\tau(n)\),
which had highest weight \(n \lambda_2\), and one obtains
\begin{equation}
  \Gamma_{hol}(E^*\otimes L^N) \cong D(0,N-n)
\end{equation}
The quantization of the bundles \(L^n\) is therefore given
by the $\mathcal{A}_N$-modules \(M^N_{L^n}\)
\begin{equation}
  M^N_{L^n}=\mathrm{Hom}_\cz(D(0,N-n),D(0,N)).
\end{equation}
An easy calculation shows that as $G$-modules we have
\begin{equation}
 M^N_{L^n} \cong
 \bigoplus_{k=0}^{N-n} D(k,k+n) \quad \mathrm{for} \quad n \geq 0
\end{equation}
and
\begin{equation}
 M^N_{L^n} \cong
 \bigoplus_{k=0}^{N} D(k-n,k) \quad \mathrm{for} \quad n < 0
\end{equation}

\subsection{The Bundles $E=L^n \otimes H^{-1}$}

The isotropy representation of \(L^n \otimes H^{-1}\) is
\(\tau(n) \otimes \chi^*\) with highest weight \(\lambda_1+n\lambda_2\).
This yields
\begin{equation}
 \Gamma_{hol}(E^*\otimes L^N) \cong D(1,N-n-1).
\end{equation}
This gives as a quantization of \(L^n \otimes H^{-1}\)
the modules
\begin{equation}
 M^N_{L^n \otimes H^{-1}}=\mathrm{Hom}_\cz(D(1,N-n-1),D(0,N)).
\end{equation}
Again as $G$-modules
\begin{equation}
 M^N_{L^n \otimes H^{-1}} \cong
 \bigoplus_{k=0}^{N-n-1} ( D(k,k+n+2) \oplus D(k+1,k+n) )
 \quad \mathrm{for}
 \quad n \geq 0
\end{equation}
and
\begin{equation}
 M^N_{L^n \otimes H^{-1}} \cong
 (\bigoplus_{k=0}^{N} D(k-n-1,k+1)) \oplus
 (\bigoplus_{k=0}^{N-1} D(k-n+1,k))
 \quad \mathrm{for}
 \quad n < 0
\end{equation}

\section{Quantized \spc-Bundles and a Dirac Operator}

In the classical case we had a family of \spc-bundles
\begin{equation}
 S_m=S_m^+ \oplus S_m^-
\end{equation}
over $M$, where
\begin{eqnarray}
  S_m^+=L^m \oplus L^{m+3}\\
  S_m^-=L^{m+1} \otimes H^{-1}.\nonumber
\end{eqnarray}
Furthermore we had the twisted Dolbeault complex
\begin{equation}
  \begin{CD}
    0 @>>> \Gamma(L^m) @> \overline \partial >> \Gamma(L^{m+1} \otimes H^{-1})
    @> \overline \partial >> \Gamma(L^{m+3}) @>>> 0
  \end{CD}
\end{equation}
which we used to define the Dirac operator
\begin{equation}
 D_m=\sqrt{2}(\overline \partial + \overline \partial^*)=
 \left( \begin{array}{cc}
    0 & \ti D_m^*\\
    \ti D_m & 0 \\
\end{array} \right)
\end{equation}
where \(\ti D_m : \Gamma(S_m^+) \to \Gamma(S_m^-)\) was a G-equivariant
map fulfilling \(\ti D_m^* \ti D_m = \Delta_G - 2m -\frac{2}{3} m^2\).\\
Let \(T\) be the quantization maps defined earlier.
We define the \spc-bundle in the quantum case to be the module
\begin{eqnarray} \label{spin1}
 \tilde M_{S_m}^N=\tilde M_{S_m^+}^N \oplus \tilde M_{S_m^-}^N \nonumber\\
 \tilde M_{S_m^+}^N=M_{L^m}^N \oplus M_{L^{m+3}}^N\\
 \tilde M_{S_m^-}^N=M_{L^{m+1} \otimes H^{-1}}^N .\nonumber
\end{eqnarray}

In the following we consider the case \(N \geq \mathrm{max}(2,m+3)\).

\begin{theorem}
 There is a unique linear map
 \begin{displaymath}
  \overline \partial_N: \quad M_{S_m}^N \to M_{S_m}^N 
 \end{displaymath}
 such that the following diagram commutes
 \begin{displaymath}
   \begin{CD}
    0 @> \overline \partial>> \Gamma(L^m) @> \overline \partial >> \Gamma(L^{m+1} \otimes H^{-1})
    @> \overline \partial >> \Gamma(L^{m+3}) @> \overline \partial >> 0\\
      @.    @VV T_{L^m}^N V @VV T_{L^{m+1} \otimes H^{-1}}^N V @VV
    T_{L^{m+3}}^N V   @.\\
     0 @>\overline \partial_N>> M_{L^m}^N @> \overline \partial_N >> M_{L^{m+1} \otimes H^{-1}}^N
    @> \overline \partial_N >> M_{L^{m+3}}^N @>\overline \partial_N>> 0\\
  \end{CD}
 \end{displaymath}
 
 \(\overline \partial_N\) is equivariant and satisfies \(\overline \partial_N^2=0\).
\end{theorem}

Proof: Uniqueness is guaranteed by the surjectivity of the maps \(T\).
For existence it is enough to show that
\begin{eqnarray}
  \overline \partial (\mathrm{ker}(T_{L^m}^N)) \subset
  \mathrm{ker}(T_{L^{m+1} \otimes H^{-1}}^N)\nonumber \\
   \overline \partial (\mathrm{ker}(T_{L^{m+1} \otimes H^{-1}}^N)) \subset
  \mathrm{ker}(T_{L^{m+3}})\nonumber
\end{eqnarray}
This is easily derived from the decomposition of the modules and the fact
that \(\overline \partial\) and the quantization maps are
equivariant. \hfill \(\Box\)

Explicitely the map is given by \(\overline \partial_N f =
T^N \overline \partial (T^N)^{-1}(f)\), where \((T^N)^{-1}(f)\)
is an arbitrary element of the set \((T^N)^{-1}(\lbrace f \rbrace)\).

\begin{definition}
  The \spc-Dirac operator \(D_m: M_{S_m}^N \to M_{S_m}^N\) is
  the self-adjoint operator, given by
  \begin{displaymath}
    D_m = \sqrt{2} (\overline \partial_N + \overline \partial_N^*)
  \end{displaymath}
\end{definition}

As in the classical case we have
\begin{displaymath}
 D_m=
 \left( \begin{array}{cc}
    0 & \ti D_m^*\\
    \ti D_m & 0 \\
\end{array} \right)
\end{displaymath}
and the map \(\tilde D_m: M_{S_m^+}^N \to M_{S_m^-}^N\) is equivariant.
We clearly have

\begin{lem}
   \item \(\tilde D_m^* \tilde D_m\vert_{\mathrm{ker}(\tilde D_m)^\perp} =
     (\Delta -2m -\frac{2}{3} m^2)\vert_{\mathrm{ker}(\tilde D_m)^\perp}\)
\end{lem}

which gives, using the decomposition of the G-modules \(M_{S_m^+}^N\) and
\(M_{S_m^-}^N\):

\begin{theorem}
  The eigenvalues \(\lambda\) of \(D_m\) and their multiplicities \(\mathrm{mult}_\lambda\)
  are:
\begin{enumerate}
  \item \(\lambda_n^\pm=\pm \sqrt{2n(n+2)+2\vert m\vert(n+1)-2m}\)\\
    \(\mathrm{mult}_{\lambda_n^\pm}=(n+\frac{\vert m \vert +2}{2})(n+\vert m
    \vert +1)(n+1)\)\\
    \(n \in \lbrace 1,\ldots, N-m-1\rbrace \quad \textrm{if} \quad m \geq
    0\)\\
    \(n \in \lbrace 0,\ldots,N-1\rbrace \quad \textrm{if} \quad m < 0\)

  \item \(\lambda_n^\pm=\pm \sqrt{2n(n+2)+2\vert m+3\vert(n+1)+2(m+3)}\)\\
    \(\mathrm{mult}_{\lambda_n^\pm}=(n+\frac{\vert m+3 \vert +2}{2})(n+\vert m+3
    \vert +1)(n+1)\)\\
     \(n \in \lbrace 0,\ldots, N-m-3\rbrace \quad \textrm{if} \quad m \geq
    -2\)\\
    \(n \in \lbrace 1,\ldots,N\rbrace \quad \textrm{if} \quad m < -2\)
    
  \item \(\lambda=0\)\\
        \(\mathrm{mult}_{\lambda}=\frac{N+2}{2}(m^2-3m N+2N^2+2N+2)\)
        
\end{enumerate}
\(\mathrm{index}(D_m)=\mathrm{dim}(\mathrm{ker}(D_m))-\mathrm{dim}(\mathrm{ker}(D_m^*))=
\frac{3}{2}(N^2+3 N +2)\).
\end{theorem}

\begin{rem}
  The index and the number of zero modes are different from the
  commutative case. That is because the maps \(\overline \partial_N\)
  vanish also on higher representations.
  Apart from this the spectrum is truncated, as we expected.
\end{rem} 

The Dirac operator defines for each \(m \in \nz\) and
 \(N \geq \textrm{max}(2,m+3)\) an even spectral triple
 \((\mathcal{A},\mathcal{H},D)\) with
 \begin{eqnarray}
   \mathcal{A}=\mathcal{A}_N \nonumber\\
   \mathcal{H}=M_{S_m}^N \nonumber\\
   D=D_m \nonumber
 \end{eqnarray}
which gives a differential calculus over the algebra $\mathcal{A}_N$.

\subsection{A different choice of bundle} Taking for the space of 
spinor fields instead of (\ref{spin1}) the space
\begin{eqnarray}
 M_{S_m}^N=M_{S_m^+}^N \oplus M_{S_m^-}^{N} \nonumber\\
 M_{S_m^+}^N=M_{L^m}^N \oplus M_{L^{m+3}}^{N+2}\\
 M_{S_m^-}^N=M_{L^{m+1} \otimes H^{-1}}^{N+1} .\nonumber
\end{eqnarray}
we can use the same procedure to define a Dirac operator,
which except for a spectral cutoff has the same spectrum as
the classical Dirac operator. In particular the index and zero
modes are classical. In this case we lose the $\mathcal{A}_N$-module
property, the space is however a module over the algebra
$\mathcal{A}_N \oplus \mathcal{A}_{N+2} \oplus \mathcal{A}_{N+1}$.
This will be investigated elsewhere.

\section{Conclusions and Outlook}

The tools provided in this paper make it possible
to investigate scalar QFT on the fuzzy $\cp$, which
is free of any divergencies, since the algebra of functions
is finite dimensional and allows to construct a well defined
functional integral.
The attempt to construct spinors led to similar difficulties
as for the known case of the fuzzy sphere.
There the use of supersymmetry provided a solution (\cite{Gross2,Reit}).
It seems promising to use a similar extension for the $\cp$.
Noncommutative generalizations of the Dolbeault operators might
as well lead to structures proposed in ~\cite{Froe}.

\section{Acknowledgements}

The authors would like to thank W. Bulla, H. Miglbauer, P. Presnajder, 
J. P\"{u}ngel, G. Reiter for helpful discussion.

\end{document}